\documentclass[11pt,a4paper]{article}

\usepackage[T1]{fontenc}
\usepackage[utf8]{inputenc}
\usepackage{lmodern}
\usepackage{microtype}
\usepackage{textcomp}
\usepackage[a4paper,margin=1.3in]{geometry}

\usepackage{amsmath,amssymb}
\usepackage{enumitem}
\usepackage[hidelinks]{hyperref} 

\title{\textbf{A short technical comment on Bub's ``There is No Quantum World'' (arXiv:2512.18400v2)\\
and a brief remark on related Grangier's reply (arXiv:2512.22965v1)}}

\author{%
  Krzysztof Sienicki\thanks{Chair of Theoretical Physics of Naturally Intelligent Systems
  (\(\mathbb{NIS}\)\textsuperscript{\textcopyright}),
  Lipowa~2 / Topolowa~19, 05--807 Podkowa Le\'sna, Poland, European Union.
  E-mail: \texttt{niskrissienicki@gmail.com}.}%
}
\date{\today}

\begin{document}
\maketitle

\begin{abstract}
This note is a friendly technical check of Jeffrey Bub's \emph{There is No Quantum World} (arXiv:2512.18400v2).
I flag one unambiguous mathematical slip (a cardinality identity that implicitly assumes the Continuum
Hypothesis) and then point out a few places where the discussion of infinite tensor products, ``sectorization,''
and measurement updates would benefit from sharper wording.
Nothing here is meant as a critique of Bub's interpretive goals; the aim is simply to separate what is
mathematically forced from what depends on choices of algebra, representation, or philosophical stance.
I end with a short remark on Philippe Grangier's reply (arXiv:2512.22965v1).
\end{abstract}

\section{Introduction and scope}
Bub's paper has two main threads \cite{Bub2512}. First, it motivates a ``neo-Bohrian'' reading: the propositional
structure is non-Boolean, contexts matter, and probabilities are taken to express indefiniteness rather than
ignorance. Second, it discusses von Neumann infinite tensor products and the ``sectorization'' program (as used
by Van Den Bossche \& Grangier) as a proposed way to tame the measurement problem, and argues that it cannot
deliver definite outcomes for any \emph{finite} apparatus---hence an unavoidable classical ``cut.''

My comments below are purely technical: what is plainly wrong, and what is correct but needs a qualifier.

\section{One clear mathematical correction: cardinality and CH}
In the discussion of an $N$-qubit measuring device, Bub uses the (standard) intuition that ``more and more
qubits'' ultimately points toward an idealized infinite tensor product, and he links this to talk of
``uncountable dimensionality'' \cite{Bub2512}. It is helpful to state the underlying set-theoretic facts
explicitly.

For a finite register, the computational basis is labelled by the set of bitstrings of length $N$,
\[
\{0,1\}^N,
\qquad\text{so}\qquad
\lvert \{0,1\}^N \rvert = 2^N.
\]
In the usual infinite idealization, one instead meets the set of \emph{infinite} bitstrings,
\[
\{0,1\}^{\mathbb{N}},
\qquad\text{so}\qquad
\lvert \{0,1\}^{\mathbb{N}} \rvert = 2^{\aleph_0}.
\]
(Some authors compress this intuition into a heuristic notation like $2^N \to 2^{\aleph_0}$ as $N\to\infty$,
but the two cardinality statements above are the precise content.)

\textbf{Problem.} Bub then writes $2^{\aleph_0}=\aleph_1$ \cite{Bub2512}. The equality
$2^{\aleph_0}=\aleph_1$ is not a theorem of standard set theory (ZFC); it is equivalent to the Continuum
Hypothesis (CH), which is independent of ZFC \cite{GodelCH,CohenCH}. Without assuming CH, the correct statement
is simply
\[
2^{\aleph_0}=\mathfrak{c}=\lvert \mathbb{R} \rvert,
\]
and if one \emph{wants} to identify $2^{\aleph_0}$ with $\aleph_1$ one should say explicitly ``assuming CH.''

This is the only unambiguous ZFC-level slip I noticed; everything else below is best read as a request for
short qualifiers that prevent technical over-interpretation.

\section{A few technical clarifications (small qualifiers that matter)}
Several places in \cite{Bub2512} would read more cleanly with short qualifiers. The points below are not meant
as deep objections; they are mostly ``precision patches.''

\subsubsection*{(i) Infinite tensor products are not just ``limits of dimensions''}
The paper can be read as if ``the dimension becomes uncountable'' is what drives the appearance of sectors.
Even ignoring CH, that is not the best conceptual emphasis. What actually does the work in the operator-algebra
picture is the choice of a physically relevant observable algebra (often quasi-local) and the choice of a
representation/reference state used to build the infinite tensor product or GNS representation.
Sector/superselection structure is then tied to central decomposition and inequivalent representations, not to
``uncountable dimension'' by itself \cite{BratteliRobinson1}.

\subsubsection*{(ii) ``Superpositions across sectors represent mixtures'' is true only relative to an algebra}
Bub writes that superpositions across sectors ``represent mixtures,'' and illustrates this with a block-diagonal
mixture-like density operator \cite{Bub2512}. As written, this is easy to misread.

A vector $\lvert\Psi\rangle$ defines a \emph{pure} state on the full bounded-operator algebra $B(\mathcal{H})$:
\[
\omega_\Psi(A)=\langle\Psi|A|\Psi\rangle.
\]
What becomes ``mixture-like'' is typically the \emph{restriction} of that state to a physically relevant
subalgebra (macroscopic/context/tail observables), where interference terms vanish or are operationally
inaccessible \cite{BratteliRobinson1,Grangier2512}. In many algebraic discussions the physically relevant algebra
is \emph{not} $B(\mathcal{H})$ at all, but a smaller quasi-local or macroscopic von Neumann algebra (often with a
nontrivial center). In such settings, a vector state in a reducible representation need not be pure as a state
on that smaller algebra. This is precisely why it helps to say explicitly \emph{which} algebra is being used
when one speaks of ``mixtures'' or ``superselection.''

\subsubsection*{(iii) Measurement update: separate selective and non-selective maps}
In the measurement postulates, the text moves from ``the state jumps to one of the eigenstates'' (selective,
conditioned on an outcome) to writing a map that produces a mixture (non-selective, outcome ignored)
\cite{Bub2512}. This is a common shorthand, but here it matters because later discussion turns on whether
one is conditioning on an actual outcome.

\subsubsection*{(iv) ``Without decoherence'' can be read too strongly}
Bub suggests the sector story yields classicality ``without the usual appeal to decoherence'' \cite{Bub2512}.
This may be intended as a presentational point, but it can be read as ``without any dynamical/coarse-graining
story.'' In many concrete detector models, decoherence (environmental entanglement plus coarse-graining) is
exactly what suppresses interference in the \emph{accessible} observables and stabilizes macroscopic records.
Sector language can repackage this structurally, but the wording would benefit from a narrower claim.

\section{A brief remark on Grangier's reply}
Grangier's comment usefully pushes back on the slogan ``infinity never comes'' by reminding us that physics
routinely relies on idealizations (thermodynamic/continuum limits) and by stressing that CSM treats ``systems
within contexts'' as an explicit starting point rather than an ``emergence proof'' \cite{Grangier2512}.

Two small technical notes about the appendix in \cite{Grangier2512}:
\begin{itemize}[leftmargin=1.2em]
\item The ``tail observable commutes with locals \emph{in the sense that} ...'' phrasing conflates an existence
statement with a commutator statement. A standard tail-algebra definition avoids this ambiguity
\cite{BratteliRobinson1}.
\item The ``asymptotic frequency labels sectors'' example is a good intuition pump, but it should be flagged
clearly as model-/representation-dependent (the sector structure depends on the chosen representation/state
and the observable algebra).
\end{itemize}

\section{Conclusion: a short patch list}
If Bub made a few very small edits, the technical presentation would be significantly clearer:
\begin{enumerate}[leftmargin=1.2em]
\item Replace $2^{\aleph_0}=\aleph_1$ by $2^{\aleph_0}=\mathfrak{c}$ (or state ``assuming CH''), i.e.\
$2^{\aleph_0}=\mathfrak{c}=\lvert \mathbb{R} \rvert$.
\item Add a sentence noting that sectorization depends on observable algebras and representations, not merely on
``going uncountable.''
\item Qualify ``superpositions are mixtures'' as a statement about restriction to macroscopic/context algebras,
and make explicit which algebra is meant when speaking about purity/mixedness.
\item (Minor) Separate selective and non-selective measurement updates explicitly.
\end{enumerate}

\end{document}